\documentclass[aps,prd,nofootinbib,amsmath,twocolumn]
{revtex4-1}
\usepackage{amssymb,esvect,amsmath,graphicx,latexsym,amsthm,slashed,eso-pic}
\usepackage[final]{pdfpages}
\DeclareMathOperator{\ev}{eV}   \DeclareMathOperator{\gev}{GeV}    \DeclareMathOperator{\g}{g}        
     \newcommand{\cL}{{\cal L}}   \newcommand{\cO}{{\cal O}}   
 \newcommand{\vp}{\varphi} \newcommand{\half}{\frac12}
\newcommand{\ie}{{\it i.e.~}}  
    \def\g{{\bf g}}
\newcommand{\pL}{\left(} \newcommand{\pR}{\right)} \newcommand{\bL}{\left[} \newcommand{\bR}{\right]}   \newcommand{\mL}{\left|} \newcommand{\mR}{\right|}
\newcommand{\beq}{\begin{equation}} \newcommand{\eeq}{\end{equation}}
\newcommand{\bea}{\begin{eqnarray}} \newcommand{\eea}{\end{eqnarray}}
\newcommand{\alg}[1]{\begin{align} \begin{split} #1 \end{split}  \end{align}}
\newcommand{\tenx}[1]{\times 10^{#1}}
\newcommand{\Eq}[1]{Eq.~(\ref{#1})} \newcommand{\Eqs}[2]{Eqs.~(\ref{#1}) and (\ref{#2})} 
  
\newcommand{\Fig}[1]{Fig.~\ref{#1}} 
\newcommand{\Tab}[1]{Tab.~\ref{#1}}
\newcommand{\App}[1]{Appendix~\ref{#1}}
\def\lsim{\mathrel{\raise.3ex\hbox{$<$\kern-.75em\lower1ex\hbox{$\sim$}}}}
\def\gsim{\mathrel{\raise.3ex\hbox{$>$\kern-.75em\lower1ex\hbox{$\sim$}}}}

\newcommand{\vev}[1]{\langle {#1} \rangle}

\newcommand{\ord}[1]{\mathcal{O}{(#1)}}

\begin{document}

\hspace{13cm} \parbox{5cm}{FERMILAB-PUB-15-318-A YITP-SB-15-26}~\\



\title{Inflatable Dark Matter}
\author{Hooman Davoudiasl}
\affiliation{Department of Physics, Brookhaven National Laboratory, Upton, NY 11973, USA}
\author{Dan Hooper}
\affiliation{Center for Particle Astrophysics, Fermi National Accelerator Laboratory, Batavia, IL  60510}
\affiliation{Department of Astronomy and Astrophysics, The University of Chicago, Chicago, IL  60637}
\author{Samuel D. McDermott}
\affiliation{C. N. Yang Institute for Theoretical Physics, Stony Brook, NY 11794}
\date{\today}

\begin{abstract}
We describe a general scenario, dubbed ``Inflatable Dark Matter'', in which the density of dark matter particles can be reduced through a short period of late-time inflation in the early universe. The overproduction of dark matter that is predicted within many otherwise well-motivated models of new physics can be elegantly remedied within this context, without the need to tune underlying parameters or to appeal to anthropic considerations. Thermal relics that would otherwise be disfavored can easily be accommodated within this class of scenarios, including dark matter candidates that are very heavy or very light. Furthermore, the non-thermal abundance of GUT or Planck scale axions can be brought to acceptable levels, without invoking anthropic tuning of initial conditions. A period of late-time inflation could have occurred over a wide range of scales from $\sim$\,MeV to the weak scale or above, and could have been triggered by physics within a hidden sector, with small but not necessarily negligible couplings to the Standard Model. 

\end{abstract}

\maketitle

\section{Introduction and Motivation}

A variety of well-established astrophysical and cosmological observations support the conclusion that there exists a form of matter whose interactions with the particles of the Standard Model
have thus far eluded detection.  The evidence for this mysterious substance, referred to as dark matter (DM),
is strong and encompasses a wide range of astronomical distance scales, from galactic to cosmological. Although the current evidence for DM is based solely on its gravitational effects, DM is expected to experience other interactions
that govern its production in the early universe. The existence of DM has motivated a wide range of experimental efforts, designed to detect the very feeble interactions between DM and the particles of the Standard Model.

Although the DM could plausibly consist of particles with a very wide range of characteristics, weakly interacting massive particles (WIMPs) represent the most broadly studied class of DM candidates. Such candidates have been motivated in large part by the realization that a stable particle with an approximately weak-scale mass and weak-scale annihilation cross section is {\it calculated} to freeze out in the early universe with a thermal relic abundance, $\Omega_{\rm WIMP}^{\rm th}$, similar to the {\it measured} cosmological DM density, $\Omega_{\rm DM}^{\rm meas} \simeq 0.26$ \cite{Ade:2015xua}. Since such particles are also theoretically motivated by the quantum stability of the Higgs potential (the ``hierarchy problem"), this is sometimes referred to as the ``WIMP miracle''.  In recent years, however, the null results of and increasingly stringent constraints from direct detection experiments and the Large Hadron Collider (LHC) have slashed into this parameter space and consequently tempered much of the enthusiasm for the WIMP hypothesis. The sensitivity of experiments attempting to detect the elastic scattering of DM with nuclei has increased at an exponential rate over the past two decades, on average nearly doubling in reach each year. As this march toward increasingly stringent constraints has progressed, many otherwise well-motivated varieties of WIMPs have become untenable. WIMPs that remain experimentally viable are generally depleted in the early universe through processes that do not induce a large elastic scattering cross section with nuclei, such as through efficient coannihilations~\cite{Griest:1990kh}, resonant annihilations~\cite{Griest:1990kh,Hooper:2013qjx}, or annihilations to final states consisting of leptons, gauge bosons, Higgs bosons, or particles residing within a hidden sector. In this paper, we present another way to decouple the DM's elastic scattering and production cross sections from its relic abundance.

A qualitatively different but also very well-motivated DM candidate is the QCD axion, $a$ \cite{Weinberg:1977ma,Wilczek:1977pj}. This particle is a consequence of the Peccei-Quinn mechanism \cite{Peccei:1977hh,Peccei:1977ur}, which can dynamically account for the extreme smallness of the CP-violating parameter, $\bar\theta\lesssim 10^{-10}$, in strong interactions. The QCD axion is a pseudoscalar with a mass that is fully determined by a symmetry-breaking scale, $f_{\rm PQ}$. Stellar and laboratory constraints require that $m_a \lesssim 10^{-2}$ eV, corresponding to $f_{\rm PQ} \gtrsim 10^9$~GeV. The axion abundance (generated via misalignment production), $\Omega_a^{\rm th}$, scales roughly linearly with $f_{\rm PQ}$, and is close to $\Omega_{\rm DM}^{\rm meas}$ for $f_{\rm PQ}\sim 10^{12}$~GeV \cite{Preskill:1982cy,Abbott:1982af,Dine:1982ah}, corresponding to $m_a \sim \cO( \mu$eV$)$ (for a review, see Ref.~\cite{Kawasaki:2013ae}). For the most theoretically well-motivated values of $f_{\rm PQ}$\,$\sim$\,$M_{\rm GUT}\sim 10^{16}$~GeV or $M_{\rm Planck}\sim 10^{19}$~GeV \cite{Svrcek:2006yi}, the value of $\Omega_a^{\rm th}$ is predicted to be very large, overclosing the universe. Anthropic selection of the misalignment angle has been proposed as a way to evade this conclusion~\cite{Wilczek:2004cr}. The class of scenarios described in this paper provides a different, non-anthropic solution.

In light of these considerations, we are motivated to propose scenarios in which the thermal history of the early universe departs from the standard radiation-dominated picture, putting theories that seem to overproduce DM into agreement with the observed DM abundance. This is particularly attractive given the host of DM candidates that arise from well-motivated theories that represent compelling extensions of the Standard Model, apart from their DM candidates.

The central idea of this paper is that if there was a period of late-time inflation after the production of DM (thermally or otherwise), models with $\Omega_X^{\rm th} \gg \Omega_{\rm DM}^{\rm meas}$ can be viable. During the reheating phase that follows inflation, the injection of entropy dilutes the abundance of DM. Such a scenario is quite general, requires no fine-tuning, and could have taken place at any time prior to big bang nucleosynthesis (BBN).\footnote{The phenomenology of this scenario is similar, in some ways, to the dilution of DM through the late-time decays of moduli or other massive states~\cite{Fornengo:2002db,Gelmini:2006pq,Hooper:2013nia,Kane:2015jia,Patwardhan:2015kga}.}

In exact analogy to how primordial inflation mitigates the monopole problem, late-time inflation reopens theory-space windows that would otherwise be closed by particle overproduction in the early universe. For instance, a period of post-freeze-out inflation can enable very heavy DM particles ($m_X \gsim 100$ TeV) to be thermal relics, evading the otherwise very general unitarity bound on their annihilation cross section~\cite{Griest:1989wd}. Similarly, inflation after the QCD phase transition enables very light axions (arising as a result of symmetry breaking at energies from $\sim$\,$M_{\rm GUT}$ to $\sim$\,$M_{\rm Planck}$) with an order one misalignment angle to be reconciled with the observed DM density.  Because the sector responsible for this inflationary phase could be almost entirely decoupled from the Standard Model and DM sectors, we are able to discuss this general class of scenarios independently of the DM theory under consideration. A period of late-time inflation can also dilute the abundance of problematic relics, such as moduli and gravitinos~\cite{Lyth:1995ka}.

In the next two sections, we will offer a brief discussion of the general concepts underlying
late-time inflation and describe a toy model in which the main features of this scenario are illustrated. Extensions and more complicated scenarios are of course possible, but will generally not change the main points we intend to convey.

\section{Exponential Expansion at Late Times}

When the energy density of the universe is dominated by a term with negative pressure, exponential expansion occurs. Our viewpoint is that this behavior is quite generic: if {\it any} scalar potential energy dominates the energy density at {\it any time}, exponential expansion will occur. Furthermore, the field that sources this potential energy density can be sequestered from known fields and need not be coupled via renormalizable interactions to any other sector of the universe.
And as long as the late-time inflation does not occur during or after BBN, it will have no adverse effects on the well-established concordance cosmology.  


Let us define $\rho_\Lambda$ to be the sum of all temperature-independent contributions to the stress-energy tensor of the universe that are proportional to the metric \cite{Weinberg:1988cp,Bellazzini:2015wva},
\beq
\langle \Theta_{\mu \nu} \rangle = \rho_\Lambda g_{\mu \nu},
\eeq
such that $\rho_\Lambda$ is independent of the scale factor. In addition to an omnipresent cosmological constant term, $\rho_\Lambda$ receives contributions from the minima of particle potentials, including the trace anomalies of confining potentials.

The universe inflates whenever $\rho_{\Lambda}$ dominates the total energy density. In the early universe, this generally corresponds to the condition $\rho_\Lambda > \rho_R$, where $\rho_R$ is the energy density in radiation. In the current epoch, $\rho_\Lambda$ in the form of ``dark energy'' dominates the energy density of the universe, and a period of exponential expansion has recently begun as a result. There is much confusion regarding the density of dark energy; na\"ive dimensional analysis would lead us to expect a vacuum energy density that is vastly larger than is observed.

\begin{figure}[t]
\includegraphics[width=0.44\textwidth]{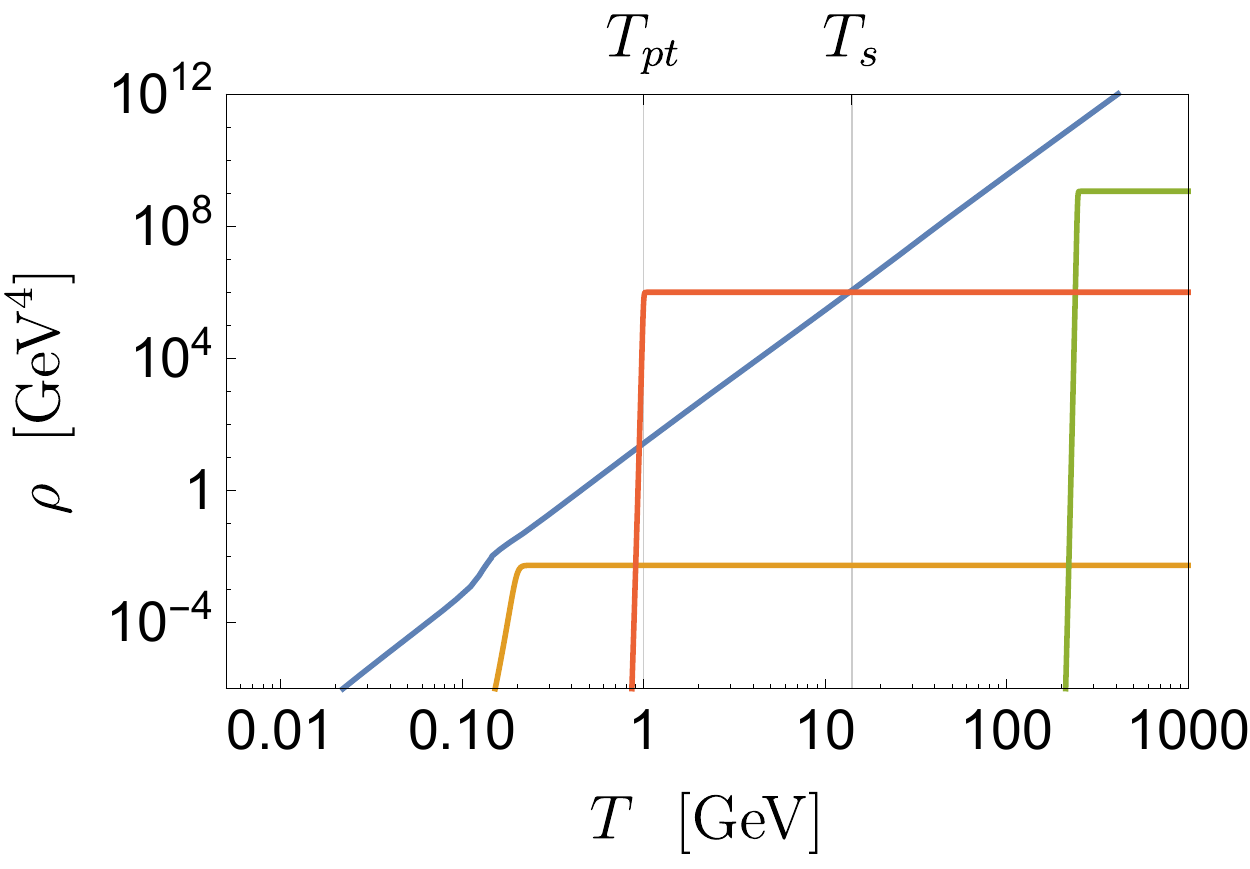}
\caption{The density of the vacuum energy associated with the electroweak (green) and QCD (orange) phase transitions, compared to the radiation energy density of the universe (blue), as a function of temperature. We also show a line for the vacuum energy density associated with a new phase transition (red), as discussed in the text. Schematically, a period of inflation occurs when the vacuum energy density exceeds the radiation energy density, starting at $T_s$ and ending at $T_{pt}$. Reheating is not fully captured by this plot.}
\label{schematic}
\end{figure}

At times in the early universe, there were other significant sources of vacuum energy, including those associated with the trace anomaly of QCD near $\Lambda_{\rm QCD}$ and the Higgs minimum near the electroweak phase transition. In each of these cases, however, the fields transitioned to their broken phase before they came to dominate the energy density of the universe, and thus did not provoke a period of late-time inflation (see, however, Refs.~\cite{Boeckel:2011yj,Schettler:2010wi,Boeckel:2009ej}). In \Fig{schematic}, we show the values of the QCD and electroweak potential energy minima as a function of temperature, and compare this to the energy density in radiation. We take the QCD phase transition temperature from Ref.~\cite{Bazavov:2014pvz}, the value of the constant part of the QCD trace anomaly from Ref.~\cite{Bazavov:2009zn}, and the number of degrees-of-freedom for the whole temperature range from Ref.~\cite{Drees:2015exa}. We simply treat the electroweak transition as a step function with coupling $\lambda_h = v_h^2/6m_h^2$, $m_h=125\gev$, and $v_h = 174 \gev$. The vacuum energy density associated with the electroweak phase transition is similar in a simple two Higgs doublet model with supersymmetric parameter relations. Contributions from matter and dark energy do not appear within the range of energy densities included in this plot. Given only the QCD and electroweak phase transitions, it seems unlikely that late-time inflation would occur. In moving forward, we will focus on inflation that is triggered by the potential energy density of fields that reside beyond the Standard Model.

An additional period of exponential expansion in the early universe could have major qualitative ramifications for the matter density of the universe. The density of any particle species that had already decoupled by the beginning of this inflation will be diluted, leading us to alter our expectations for the interaction strength, mass, and other characteristics of the DM. In the following section, we will consider a toy model to explore the phenomenology resulting from a period of late-time inflation driven by a phase transition within a hidden sector.

\section{A Toy Model}
\label{toy}

Consider a scalar field, $\phi$, for which at high temperatures $\langle \phi \rangle =0$, with a potential energy density $V_0$. At late times, the field exits this false vacuum and moves into a symmetry breaking minimum, $\phi \to v_\phi + \vp$, corresponding to a state with zero energy density. We assume that the potential energy in the early universe rests at a false minimum: the potential well that contains the field at $\langle \phi \rangle =0$ is sustained by thermal corrections to the low-energy potential, and when these are absent the global minimum occurs at $\langle \phi \rangle =v_\phi$.

The scale factor increases exponentially for as long as the scalar potential energy dominates the energy density of the universe. The inflationary phase causes the matter in the universe to cool, which removes the potential barrier and allows $\phi$ to relax into its lower energy, broken phase at $\langle \phi \rangle = v_\phi$. After a brief period of matter domination, the coherent oscillations of the newly massive scalar field will decay (as will any modes which are given mass by the spontaneous symmetry breaking), reheating the universe and restoring radiation domination.

For our toy model, we adopt a very simple zero-temperature scalar Lagrangian:
\beq \label{zero T Lagrangian}
\cL \supset - \frac12 \mu_\phi^2 \phi^2 + \sum_f y_f \bar f f \phi + \frac1{4!} \lambda_\phi \phi^4.
\eeq
Here, $\phi$ is a real scalar coupled to some number of light fermions, $f$. Because of the relative signs of the quadratic and quartic terms in \Eq{zero T Lagrangian}, $\phi$ acquires a vacuum expectation value (vev) of $v_\phi = \langle \phi \rangle = \mu_\phi \sqrt{6/\lambda_\phi}$ at low temperature. The associated Higgs particle has a mass given by $m_\vp^2 =2 \mu_\phi^2$. For ease of discussion, we assume that this sector is in thermal equilibrium with the Standard Model (which can be accomplished, for example, by a very small degree of mixing between $\phi$ and the Standard Model Higgs), but we stress that the phenomenology is easily translated to an even more sequestered sector. \Eq{zero T Lagrangian} contains all the ingredients necessary for a period of late-time inflation.

At high temperatures, the scalar seeks to minimize the following leading-order effective potential~\cite{Quiros:1999jp}:
\begin{multline} \label{thermal potential}
V(\phi) =   V_0 +  \frac1{4!} \lambda_\phi \phi^4 + \\ + \half \bL \frac1{12} \pL \sum_f \g_f y_f^2 + \half \lambda_\phi  \pR T^2 - \mu_\phi^2 \bR \phi^2   ,
\end{multline}
where $V_0=3\mu_{\phi}^4/2\lambda_\phi$ is defined such that $V(v_\phi)=0$, $\g_f=1(2)$ for a Majorana (Dirac) fermion, and the sum is over all fermions with mass $m_f \lesssim T$. A phase transition between a high-temperature, symmetric phase and a low-temperature, symmetry-broken phase occurs when the mass term in \Eq{thermal potential} passes from positive to negative, at a temperature given by:
\beq \label{Tt}
T_{pt} = \sqrt{\frac{12}{\sum_f \g_f y_f^2 + \lambda_\phi/2}} \,\,\, \mu_\phi.
\eeq
The vacuum energy density abruptly and dramatically decreases at temperatures below the phase transition.

We note that in the absence of a protective symmetry, a cubic term could have also appeared in \Eq{zero T Lagrangian}. Because of the non-analytic nature of the thermal potential, cubic terms are also generated at higher order in $\lambda_\phi$ in \Eq{thermal potential}, even in the presence of a protective symmetry~\cite{Quiros:1999jp}. A sufficiently large positive cubic term will trap the $\phi$ field at the origin until it tunnels away, potentially inducing a first order phase transition and impairing our ability to reliably calculate the predictions of the model. Thus, for ease of discussion, we (conservatively) assume in our toy model that a residual ${\mathbb Z}_2$ symmetry prohibits a cubic term in \Eq{zero T Lagrangian}, and that $\lambda_\phi$ is small enough that the cubic thermal term generated at higher order does not impact the phase transition.

Inflation starts in our toy model when $V_0$ exceeds the radiation energy density, $\rho_R(T) = \pi^2 g_{\rm eff}(T) \,T^4/30$. This occurs at a temperature given by:
\beq \label{T*}
T_s \simeq \mu_\phi \pL \frac{45}{\pi^2 g_s \lambda_\phi} \pR^{1/4}.
\eeq
Because inflation ends when the phase transition occurs, inflation is only possible if $T_s \gsim T_{pt}$. Observations of the cosmic microwave background (CMB) and the light element abundances indicate that the baryon abundance was not significantly diluted after BBN~\cite{Cyburt:2015mya}, thus requiring that $T_s \gsim T_{\rm BBN}$. We discuss other requirements for successful inflation, including the exit of the $\phi$ field from the symmetric phase, in more detail in \App{inflation}. In \Fig{schematic}, we show a schematic illustration of a model with late-time inflation.

We emphasize that late-time inflation can arise within the context of many theories, and is by no means limited to the simple scalar theory presented here. Very similar phenomenology could just as easily come about from a QCD-like gauge theory where confinement occurs at a temperature that is considerably lower than the fourth root of the trace anomaly. In this case, dedicated lattice simulations might be required to understand the physics of the phase transition.

\begin{figure}[t]
\begin{center}
\includegraphics[width=0.44\textwidth]{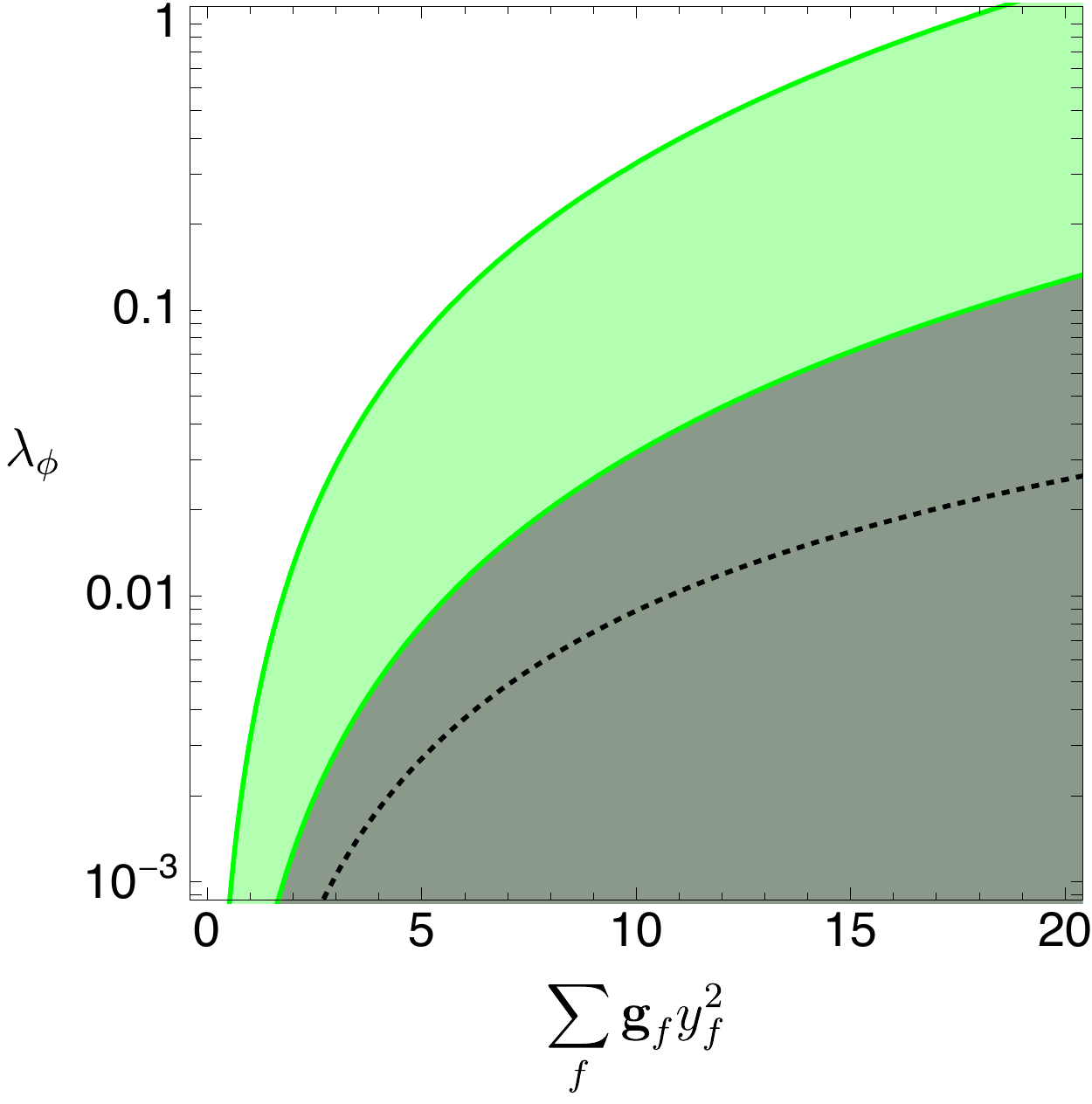}
\caption{For the toy model described by \Eq{zero T Lagrangian}, inflation occurs in the green (gray) region of the parameter space for $g_s=10~(100)$. With three Dirac fermions as the only additional matter, the toy model may be fine-tuned in the region below the dotted black line; see text for further discussion.}
\label{inflation-criterion}
\end{center}
\end{figure}

In terms of the ultraviolet (UV) couplings, the criterion for inflation, $T_s \geq T_{pt}$, is a quadratic equation in the coupling, $\lambda_\phi$. There are two critical values for inflation to be possible, at large and small values of $\lambda_\phi$. The coupling in the large-value branch is non-perturbative, and we will neglect this possibility moving forward. In the small-value branch (for which $\lambda_{\phi} \ll 2\sum_f \g_f y_f^2$), however, the critical value for inflation is given by
$\lambda_\phi \lesssim \lambda_c $, where
\beq \label{lambdac}
\lambda_c \equiv \frac5{16 \pi^2 g_s} \Big( \sum_f \g_f y_f^2 \Big)^2 \simeq \frac{N_D^2 y^4}{790}\times \bigg(\frac{100}{g_s}\bigg).
\eeq
In the second step of this equation, we have assumed that the additional particles consist of $N_D$ Dirac fermions, each with the same coupling, $y$. In \Fig{inflation-criterion}, we show the parameter space in which inflation occurs in this model, using the exact forms of \Eqs{Tt}{T*}.

Because the fermion couplings are quite a bit larger than the scalar self-coupling, $\lambda_\phi$ gets significantly renormalized by the following term: $\delta \pL \lambda_\phi \pR_f \sim  -\sum_f y_f^4/16\pi^2$~\cite{Cohen:2008nb}. We may be concerned about stability or fine-tuning when $\lambda_\phi \lesssim \mL \delta \pL \lambda_\phi \pR_f \mR$. From the model building point of view, however, the presence of fermions in \Eq{zero T Lagrangian} instead of scalars or gauge bosons is completely incidental to the discussion: new bosons that couple to $\phi$ with coupling $c_\phi$ also add $c_\phi T^2 \phi^2/48$ per degree-of-freedom to \Eq{thermal potential}, but add positively to the renormalization of $\lambda_\phi$. Because the spin of the new $\phi$ partners are immaterial in the toy model presented here, stability is not a grave concern. But in the absence of some principled cancellation (as, for instance, is possible in supersymmetric sectors) fine-tuning persists, so we point out that the region below the dotted black line in \Fig{inflation-criterion} is potentially fine-tuned, assuming $N_D=3$ Dirac fermions with equal Yukawa couplings. The parameter $\mu_\phi^2$ is also quadratically renormalized and thus sensitive to high energy scales and potentially fine-tuned. However, we point out that this naturalness problem may be resolved by the same sector that supplies a dark matter candidate, which need only be approximately an order of magnitude above the inflationary physics scale.

During inflation, the scale factor grows exponentially while the temperature and density of matter and radiation are exponentially reduced. Because the temperature is falling, the potential barrier is ``quietly'' removed after the phase transition at $T_{pt}$; the $\phi$ field will exit its low-entropy state at $\phi=0$ and inflation ends gracefully \cite{Lyth:1995ka}. As the field commences coherent oscillations, particles in the $\phi$ zero-mode are created. The universe will subsequently reheat and resume a radiation-dominated phase, allowing the standard cosmological history to ensue.

In transitioning back to radiation domination, a large entropy density, $s$, is dumped into the thermal bath. This causes a dilution of the normalized DM density. The conserved quantity that determines the observed DM relic abundance is $n/s$. Assuming that all particles are in thermal equilibrium and that there is no entropy production during inflation, we have $n_{pt}/s_{pt}=n_*/s_*$, where subscripts ``$*$'' denote the values prior to any late-time entropy production. If reheating is brief (with no matter-dominated expansion) we also have $n_{\rm RH}=n_{pt}$. Using these relations, the factor that describes the dilution of $n/s$ due to the period of inflation, $n_{\rm RH}/s_{\rm RH} \equiv \Delta^{-1} \times n_*/s_*$, is simply $\Delta=s_{\rm RH}/s_{pt}$. In terms of the UV parameters of our toy model, we find~\cite{Cohen:2008nb}:
\alg{ \label{dilution factor}
\Delta_\phi &= \frac{g_{\rm RH}}{g_{pt}} \pL \frac{T_{\rm RH}}{T_{pt}} \pR^3
 =  \bL \pL \frac{g_{\rm RH}}{g_{pt}} \pR^{1/3} + \frac{g_{\rm RH}^{1/3}g_s}{g_{pt}^{4/3}}  \frac{\lambda_c}{\lambda_\phi} \bR^{3/4},
}
where in the second step we used the conservation of energy, $\rho_R(T_{\rm RH}) = \rho_R(T_{pt}) + \rho_\phi$, with $\rho_\phi = \rho_R(T_s)$ and $\lambda_c$ as defined in \Eq{lambdac}.

\begin{figure}[t]
\begin{center}
\includegraphics[width=0.44\textwidth]{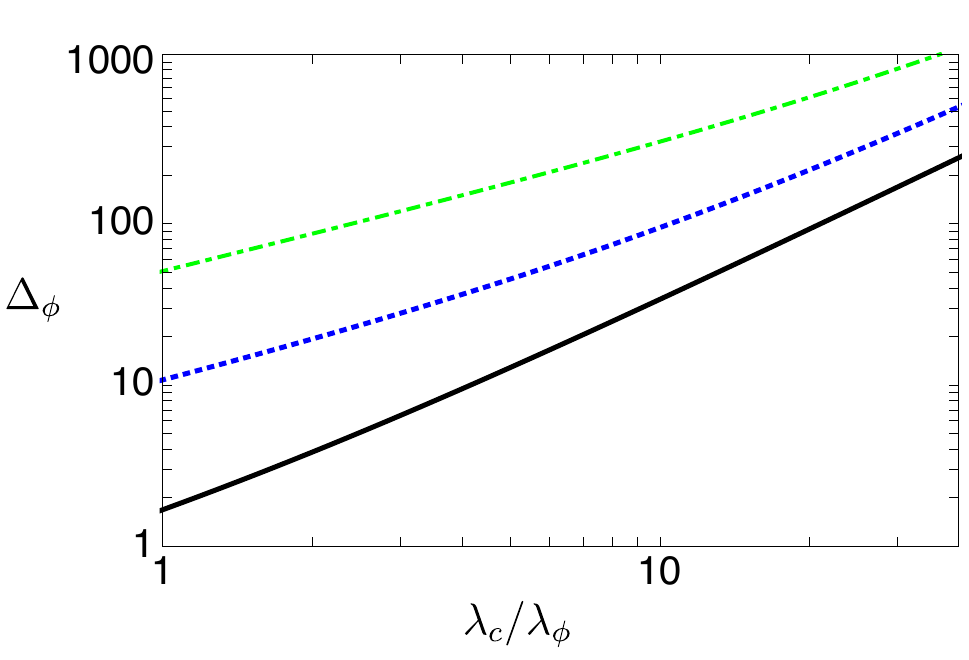}
\caption{The dilution factor in our toy model, $\Delta_\phi$, as a function of the self-coupling, $\lambda_{\phi}$, in units of the critical value needed to produce inflation. For the black solid (blue dotted) \{green dot-dashed\} line, we take $g_{\rm RH}/g_{pt}=1\,(10)\,\{50\}$ and set $g_s=g_{\rm RH}$.}
\label{dilution}
\end{center}
\end{figure}

In \Fig{dilution}, we plot the dilution factor, $\Delta_\phi$, as a function of $\lambda_c/\lambda_\phi$ for representative values of $g_{\rm RH}$ and $g_{pt}$, assuming $g_{\rm RH} = g_s$.\footnote{We note that in our model we typically expect $g_s > g_{\rm RH}$, due to the decoupling of fermions after reheating (see the Appendix for details).  Hence, our plotted dilution factor is a conservative underestimate.} These values are of course fixed for a set of UV parameters. In particular, we expect $g_{\rm RH}/g_{pt}$ to increase monotonically with $\lambda_c/\lambda_\phi$, because as $\lambda_\phi$ gets smaller the amount of exponential expansion increases, and the number of kinematically accessible degrees-of-freedom at the end of inflation decreases correspondingly. Nonetheless, the exact values of $g_{\rm eff}(T)$ depend on the temperature at which inflation starts. 

The dilution factor presented in \Eq{dilution factor} is entirely general; below, we will use $\Delta$ with no subscript to refer to the dilution factor in a more general setting. The dilution impacts all frozen-out particles and all particle asymmetries, regardless of how strongly those particles couple to the ``inflaton''. For instance, if the inflation occurred after baryogenesis, the normalized net baryon number density of the universe would also be diluted. For dilution to be effective for a purely symmetric thermal relic, the DM must have frozen out before exponential expansion begins. For a non-thermal or asymmetric particle, the generation mechanism should be set before or because of inflation. 

Note that since freeze-out typically occurs at a temperature that is at least an order of magnitude below the particle's mass, we require thermal relic DM to have acquired its mass from a mechanism that is not associated with the phase transition responsible for inflation (see Ref.~\cite{Cohen:2008nb} for a different approach to this problem). This is not unexpected, however, as operators responsible for fermion masses can be generated at any scale where chiral symmetry is broken. For example, this mass could be the result of QCD-like chiral condensation or some other spontaneous symmetry breaking at temperatures above the weak scale. This flexibility only eases inflatable DM model building, and allows application of this framework to almost any model with overabundant DM. Even dramatic disparities in the observed and inferred relic densities can be corrected with only a brief period of exponential expansion.

\section{Implications For Dark Matter}

As discussed in the introduction, the dilution from late-time inflation can resuscitate theoretically attractive DM models that would be ruled out by direct detection experiments or by the LHC under standard cosmological assumptions. In this section, we discuss some of the implications of late-time inflation for DM model building.

\subsection{Thermal Relic Dark Matter Candidates}
\label{thermal}

\begin{figure*}[t]
\begin{center}
\includegraphics[width=0.69\textwidth]{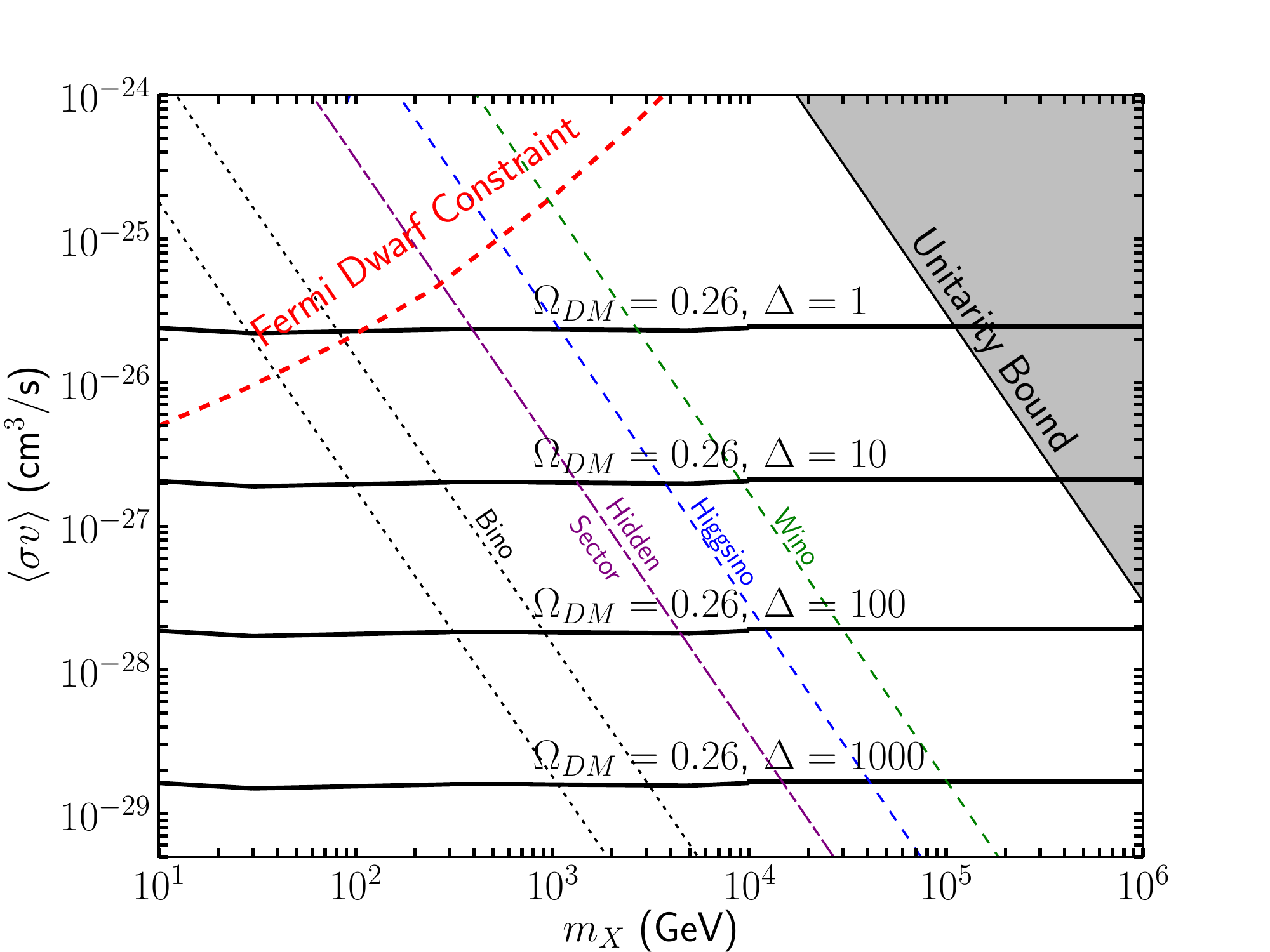}
\caption{The effective annihilation cross section at freeze-out for four benchmark DM models (described in the text). The approximately horizontal thick black lines represent the values of the cross section that generate the observed dark matter density, for four different values of the dilution factor, $\Delta$. Large low-velocity annihilation cross sections, as shown in the upper left region, are ruled out by Fermi's observations of dwarf galaxies (assuming that the annihilation cross section is the same at low velocities and freeze-out)~\cite{Ackermann:2015zua}. The upper right region is incompatible with partial wave unitarity~\cite{Griest:1989wd}.}
\label{sigmav}
\end{center}
\end{figure*}

The cosmological abundance of DM candidates that were in thermal equilibrium in the early universe is determined by their ability to deplete their number density via self-annihilation. For thermal relics, the surviving abundance (under standard cosmological assumptions) scales as $\Omega_X^{\rm th} \propto \langle \sigma v \rangle^{-1}$, where $\langle \sigma v \rangle$ is the thermally averaged annihilation cross section of the DM candidate. Over a wide range of masses, DM candidates with an annihilation cross section of approximately $\langle \sigma v \rangle_{\rm th} \sim 2 \times 10^{-26}$ cm$^3$/s freeze-out with a thermal relic abundance in agreement with the measured cosmological DM density; higher or lower cross sections lead to a DM abundance that is too small or too large, respectively. Candidates with larger cross sections generally provide only a sub-dominant fraction of the DM, and must be supplemented with another source or sources of DM. While perhaps a departure from minimality, such a scenario does not pose any phenomenological problems. Candidates with smaller cross sections, in contrast, lead to the overproduction of DM. Any theory that contains a stable thermal relic with annihilation cross section $\langle \sigma v \rangle \lesssim \langle \sigma v \rangle_{\rm th}$ (evaluated at freeze-out) is ruled out under standard cosmological assumptions.

To be concrete, we will consider the following set of thermal DM benchmark models:
\begin{itemize}
\item{{\bf A pure bino neutralino}: Pure or nearly pure binos annihilate largely through the $t$-channel exchange of sfermions. If the sleptons are significantly lighter than the squarks, the right-handed charged sleptons will be the most efficient mediator. Results from the LHC and LEP constrain the slepton masses, reducing the possible bino annihilation cross section and thus increasing its minimal relic abundance.}
\item{{\bf A pure higgsino neutralino}: The supersymmetric parameter, $\mu$, sets the masses of the two neutral and two charged higgsinos, leading to four quasi-degenerate states that each play a role in the process of thermal freeze-out. Higgsinos annihilate and coannihilate largely to gauge boson pairs, through the $t$-channel exchange of the four higgsino states. A modest enhancement in the effective annihilation rate at freeze-out results from Sommerfeld enhancements.}
\item{{\bf A pure wino neutralino}: The wino mass parameter, $M_2$, sets the masses of the neutral and charged winos, each of which play a role in the process of thermal freeze-out. Similar to higgsinos, winos annihilate and coannihilate (largely to gauge boson pairs) through the $t$-channel exchange of the charged and neutral winos. Sommerfeld enhancements significantly increase the effective annihilation rate at freeze-out.}
\item{{\bf A simple hidden sector model}. In this model, we take the DM to be a Dirac fermion that annihilates into a pair of light gauge bosons, each of which is a Standard Model gauge singlet.}
\end{itemize}

In Fig.~\ref{sigmav}, we plot the effective annihilation cross section as evaluated at thermal freeze-out for each of these four candidates. For each of the three supersymmetric examples, we use the approximate expressions given in Ref.~\cite{ArkaniHamed:2006mb}, supplemented in the higgsino and wino cases by a factor intended to account for the effects of Sommerfeld enhancements~\cite{Hisano:2006nn}. For higgsinos and winos, the effective cross sections include the effects of coannihilations on the thermal relic abundance. For the hidden sector model, we adopt the following annihilation cross section:
\beq \label{unitarity sigma}
\langle \sigma v \rangle \sim 2 \tenx{-26}{\rm cm}^3/{\rm s} \times  \bigg(\frac{g_X}{0.35}\bigg)^4 \bigg(\frac{400\,{\rm GeV}}{m_X}\bigg)^2,
\eeq
where $m_X$ and $g_X$ are the DM mass and gauge coupling, respectively.  For the bino case in Fig.~\ref{sigmav}, the two black dotted lines denote the cross section for $m_{\tilde{l}_R}/m_{\chi} = 1.2$ and 3, where $m_{\tilde{l}_R}$ and $m_{\chi}$ are the masses of the right-handed sleptons and bino, respectively. For the hidden sector model, we have adopted a value of $g_X=0.35$ for the gauge coupling.

For the four benchmark models shown in this figure, we see that under standard cosmological assumptions ($\Delta=1$), the desired thermal relic abundance is obtained for DM with masses between approximately 30 GeV and 3 TeV.  This is roughly the mass range generally associated with the WIMP paradigm. For a similar model with larger couplings, the preferred mass range shifts upwards. This has a firm upper limit, however, because perturbation theory eventually breaks down when the coupling gets too large. A model-independent upper limit can be placed by requiring that the DM annihilation cross section respects partial-wave unitarity~\cite{Griest:1989wd}. This requirement implies $\sigma v \lsim 3 \times 10^{-22}$ cm$^3$/s $\times \, (2J+1) \, ({\rm TeV}/m_X)^2$, or $m_X \lsim 120$ TeV $\times \sqrt{2J+1}$, where $J$ is the partial wave through which annihilation occurs.

Very different conclusions are possible if DM is inflatable, however. For significant values of the dilution factor, even nearly inert particles can be perfectly viable DM candidates. In particular, thermal DM particles could be much heavier than would otherwise be possible, as indicated by
the contours in Fig.~\ref{sigmav}.  Here, we see that with a moderate amount of dilution, $\Delta$, DM candidates with masses of $\ord{10-100}$~TeV or more
can lead to the observed DM abundance, $\Omega_{\rm DM}^{\rm meas} = 0.26$.

If the experiments at the LHC do not provide any signals of new physics over the next few years,
one may be compelled to assume that there is at least a modest hierarchy between the weak scale and scale of new physics.  Apart from
an apparent fine-tuning of the weak scale, one could argue that such a scenario would lead to unacceptable
relic abundances of DM within well-studied and theoretically interesting new physics models \cite{Betre:2014fva}.  If there were a period of late-time inflation, however, phenomenological relic abundance arguments
against high scale physics would no longer pose serious problems, potentially opening a new landscape of possibilities for model building.

We also add that if the DM has a mass below a few GeV, stringent constraints from observations of the CMB may imply that it cannot be a thermal relic (unless strongly $p$-wave suppressed)~\cite{Madhavacheril:2013cna}. If the DM is inflatable, however, it could have a sub-thermal cross section and still constitute the observed DM density.  For a $\sim 1$~GeV DM particle, the current constraint (the cosmic variance limit) from the CMB power spectrum is $\langle \sigma v \rangle /  \langle \sigma v \rangle_{\rm th} \lsim 0.2$ (0.02) \cite{Madhavacheril:2013cna}. Inflatable DM may not be a feasible option for substantially lower masses, as the freeze-out temperature gets too close to that of the BBN era, $T_{\rm BBN}\sim 1$~MeV.

\subsection{The QCD Axion}

The axion, $a$, is a well-motivated non-thermal DM candidate~\cite{Weinberg:1977ma,Wilczek:1977pj} that arises dynamically from the Peccei-Quinn solution to the strong CP problem~\cite{Peccei:1977hh,Peccei:1977ur}. The axion energy density is roughly proportional to a heavy mass scale, $f_{\rm PQ}$, and the axion mass and couplings are proportional to $f^{-1}_{\rm PQ}$. The QCD axion field only acquires a mass after the QCD phase transition, when it begins to coherently oscillate. Neglecting contributions from the decay of topological defects, the abundance of QCD axions scales as $\Omega_a^{\rm th} \propto  f_{\rm PQ}^{1.175} \theta_m^2$, where $\theta_m$ is the misalignment angle~\cite{Agashe:2014kda}. Despite the fact that axions are constrained to be very light ($m_a \lesssim 10^{-2} \ev$), they behave like cold DM as a result of their particular production mechanism and due to the fact that they do not thermalize with the rest of the universe.

For ``natural" values of the misalignment angle, $\theta_m\sim \ord1$, the axion will be produced with an abundance equal to $\Omega_{\rm DM}^{\rm meas}$ for $f_{\rm PQ} \sim 10^{12}$~GeV \cite{Preskill:1982cy,Abbott:1982af,Dine:1982ah}. For higher values of the Peccei-Quinn scale, such as those motivated by string theory compactifications with $f_{\rm PQ} \sim 10^{16}$~GeV~\cite{Svrcek:2006yi}, the axion abundance will dramatically exceed $\Omega_{\rm DM}^{\rm meas}$ unless the misalignment angle is tuned to very small values, perhaps as a result of anthropic selection~\cite{Wilczek:2004cr}. Alternatively, a period of late-time inflation could dilute the axion abundance, bringing it into accordance with $\Omega_{\rm DM}^{\rm meas}$.

In order to alter the cosmological abundance of axions, a period of late-time inflation would have to occur after the axion acquires its mass during the QCD phase transition. Hence, if a QCD axion is to be an untuned remnant of string dynamics at their natural scale, a period of inflation at a temperature of $\mathcal{O}{(1-100)}$ MeV is required. This scenario of ``misanthropic misalignment'', with untuned high-scale axions, represents a non-anthropic alternative to reconcile the theoretical preference for $f_{\rm PQ} \gg 10^{12}$~GeV with the measured abundance of dark matter.

\subsection{Asymmetric Dark Matter}

If there exists a primordial asymmetry between the matter and antimatter components of the DM, $\eta_X \equiv (n_X - n_{\bar X})/s$, the process of thermal freeze-out can be qualitatively altered. In asymmetric DM models, annihilations cease when either the $X$ or $\bar{X}$ population is almost entirely depleted, in contrast to symmetric models in which freeze-out occurs when Hubble expansion comes to dominate over the annihilation rate. Fixing the DM asymmetry to the baryon asymmetry provides a potential solution to the coincidence problem (\ie that $\Omega_{\rm DM} \sim \Omega_b$) and furnishes a DM candidate at light scales~\cite{Kaplan:2009ag} without relying on the standard process of thermal freeze-out to set its abundance (see Refs.~\cite{Davoudiasl:2012uw,Petraki:2013wwa,Zurek:2013wia} and references therein).  Investigations into the relation between the magnitude of the asymmetry and the requirements on the annihilation cross section have revealed a continuum of asymmetric WIMP DM models \cite{Graesser:2011wi,Iminniyaz:2011yp,Lin:2011gj}.

\begin{table}[t]
\begin{center}
\text{Standard Cosmology}~\\~\\
\begin{tabular}{cccl}
Symmetric DM:&$\langle \sigma v \rangle \simeq \langle \sigma v \rangle_{\rm th},$ & $\eta_X^{\rm i} \ll \eta_B^{\rm meas}$  \\~\\
Asymmetric DM:& $\langle \sigma v \rangle \gg \langle \sigma v \rangle_{\rm th},$ &  $\eta_X^{\rm i} \simeq \eta_B^{\rm meas}$
\end{tabular}~\\~\\~\\ \hrule~\\
\text{Cosmology With Late-Time Inflation}~\\~\\
\begin{tabular}{cccl}
Symmetric DM:&$\langle \sigma v \rangle \ll \langle \sigma v \rangle_{\rm th},$ & $\eta_X^{\rm i} \ll \eta_B^{\rm meas}$  \\~\\
Asymmetric DM:& $\langle \sigma v \rangle \gg \langle \sigma v \rangle_{\rm th},$ &  $\eta_X^{\rm i},\eta_B^{\rm i} \gg \eta_B^{\rm meas}$
\end{tabular}
\end{center}
\caption{Representative examples of scenarios with and without a period of late-time inflation that can lead to an acceptable DM abundance. In the case of non-self-conjugate DM with a sizable primordial asymmetry, many phenomenological possibilities exist (see text for discussion). The superscript ``i" refers to initial asymmetries.}
\label{IADM}
\end{table}%

Because of the porous boundary between asymmetric and symmetric DM, a period of late-time inflation can have a wide range of effects on this class of DM models.  The nature of the DM population that exists after a period of late-time inflation depends on the magnitudes of the initial asymmetry and annihilation cross section, and on whether the process of reheating produces equal numbers of matter and antimatter particles (\ie whether reheating is symmetric). If the reheating temperature is below the temperature of freeze-out, the resulting relic abundance is simply diluted as in the case of a symmetric thermal relic. In the case of asymmetric dark matter, however, we can also consider scenarios in which the universe is reheated after inflation to a temperature above the freeze-out temperature, but below the (presumably much higher) temperature at which the asymmetry was initially established. For example, suppose that DM annihilation is more efficient than thermal, $\langle \sigma v \rangle \gg \langle \sigma v \rangle_{\rm th}$, and the particle asymmetry is very large, $\eta_X^{\rm i},\eta_B^{\rm i} \gg \eta_B^{\rm meas}$. Under standard cosmological assumptions, the DM would be overly abundant in this scenario. A period of late-time inflation prior to freeze-out, followed by symmetric reheating would reduce the effective asymmetry, allowing for more efficient annihilations during freeze-out and for a lower DM abundance. 

One could imagine very different scenarios if the process of reheating is itself not symmetric. For example, if the annihilation cross section is approximately thermal or somewhat smaller, $\langle \sigma v \rangle \lsim \langle \sigma v \rangle_{\rm th}$, both matter and anti-matter components of the DM population will survive the early universe, even in the presence of a significant asymmetry, $\eta_X^{\rm i},\eta_B^{\rm i} \gg \eta_B^{\rm meas}$. If pre-freeze-out inflation and reheating then occurred symmetrically, the surviving DM abundance would be too large (since the DM-symmetric population alone would saturate the observed relic density). If the reheating following inflation were instead anti-symmetric (\ie preferentially generating anti-DM over DM), however, this could restore the DM to an approximately symmetric state, allowing it to annihilate more efficiently during freeze-out. While there are many other variations we could consider, the above examples
suffice to illustrate that inflatable DM has non-trivial implications for the asymmetric DM framework.
We schematically display the ramifications of late-time inflation on some simple symmetric and asymmetric DM scenarios in \Tab{IADM}.

\section{Associated Phenomenology}

The inflatable DM framework described in this paper could take many forms, and it would be impractical to discuss the signals and consequences in all possible cases. In some scenarios, inflatable DM may not lead to signals that are easily accessible in planned experiments. In others, however, experimental signatures in support of this framework could very plausibly appear. In this section, we will consider a few representative examples that illustrate the scope (and fecundity) of the phenomenology associated with inflatable DM.

Generally speaking, observations that appear to imply DM parameter values outside of the expected range could be interpreted as a hint in favor of a non-standard cosmological history.  The detection of DM with an annihilation cross section that is too small to obtain an acceptable relic abundance is one such example. The detection of a GeV-scale WIMP-like DM particle could also be suggestive of a non-standard thermal history, since the annihilation cross section of such a candidate is already strongly constrained by CMB observations, as mentioned earlier. Although GeV-scale DM is difficult
to detect via nuclear recoils, there have been a number of proposals to probe such
particles in accelerator fixed target experiments, making use of their ``dark" sector gauge interactions \cite{Alekhin:2015byh,Batell:2009di,Dharmapalan:2012xp,Izaguirre:2013uxa,Batell:2014mga} (see also Refs.~\cite{Essig:2011nj,Essig:2012yx} for alternative approaches based on DM-electron scattering).

Another possibility, discussed earlier in this paper, that would support a scenario with late-time inflation would be the discovery of an ultra-light axion, corresponding to $f_{\rm PQ} \gg 10^{12}$~GeV. As such ultra-light axions would generally be predicted to yield a DM abundance in significant excess of $\Omega_{\rm DM}^{\rm meas}$, such a discovery would imply a departure from the standard axion DM picture, such as an anthropic tuning of the misalignment angle, or a non-standard cosmological history. Proposals for detecting
time varying CP-odd nuclear moments (such as electric dipole moments) induced by such ultra-light DM axions promise to probe this parameter space in the coming years \cite{Budker:2013hfa}.  As the mass of the axion is acquired during the QCD phase transition, the dilution of this DM candidate requires a period of late-time inflation at an energy scale of $\sim 1-100$~MeV. A hidden sector that triggers a phase transition at such a low energy scale could potentially be probed by a variety of intensity frontier experiments, even if quite weakly coupled to the Standard Model \cite{Essig:2013lka}.

Extensions of the electroweak sector, as may be required to explain the Higgs potential, provide motivation for DM
masses in the range of $\ord{10^2-10^3}~\text{GeV}$.  DM in this mass range generally freezes out at temperatures
of $\ord{10-100}~\text{GeV}$ or less.  In this case, the post-freeze-out inflation would occur at or below the $\sim$100 GeV scale. On the other hand, we argued in Sec.~\ref{thermal} that very heavy ($\sim$\,$10-100$ TeV or heavier) relics could easily yield an acceptable abundance within the context of inflatable DM. In such a scenario, inflation could occur at a relatively high temperature, around or above the TeV-scale. If the corresponding phase transition is first order, we
could expect associated gravitational wave signals to be potentially observable \cite{Grojean:2006bp}.

Thus far, we have not commented on any potential UV completions of the inflationary sector. Such a sector could plausibly originate from non-trivial low-scale dynamics in ``Hidden Valley" models \cite{Strassler:2006im} or a ``dark QCD" sector, such as arises within the twin Higgs model \cite{Chacko:2005pe} and its extensions.  This would
eliminate the introduction of further hierarchies associated with low-mass scalar inflatons.
The observation of any signals of such dark dynamics \cite{Han:2007ae,Craig:2015pha}, possibly connected to the dynamics of the dark matter itself \cite{Garcia:2015loa,Garcia:2015toa,Craig:2015xla,Farina:2015uea},
could shed light on their contribution to a period of late-time inflation.

The connections between inflatable DM and baryogenesis are also intriguing. As noted above, the dilution of relic abundances and particle asymmetries is a universal prediction of scenarios with a period of late-time exponential expansion. If evidence is found for late-time inflation (after the establishment of the baryon asymmetry), it would imply that the primordial baryon asymmetry must have been much larger (by a factor of $\Delta$) than the value implied under standard cosmological assumptions by observations of the CMB and the light element abundances. It would also be intriguing to consider the possibility that a baryon asymmetry could be generated through the process of late-time reheating.

\section{Conclusions}

In this paper, we have explored a generic mechanism for reducing the abundance of dark matter (DM) in the early universe, allowing us to bring theories that predict unacceptably high DM densities into agreement with observations. This is accomplished through a brief period of exponential expansion (\ie inflation) taking place at late times, but prior to big bang nucleosynthesis. Such late-time inflation is quite generic, and is predicted to occur whenever any scalar potential dominates the energy density of the universe. The vacuum energy associated with the QCD trace anomaly and with the Higgs potential each contributed significantly to the energy density of the universe in the moments leading up to the QCD and electroweak phrase transitions, respectively, but likely did not come to dominate over the density in radiation (see Fig.~\ref{schematic}). The vacuum energy density associated with another phase transition with slightly different characteristics, perhaps sequestered from the Standard Model in a hidden sector, could easily have come to dominate the energy density of the universe, bringing forth a brief inflationary era.

Within the context of a simple representative model (see Sec.~\ref{toy}), we derived the conditions for late-time inflation to occur, finding only that it requires the scalar to have a somewhat weak self-coupling, delaying the onset of the corresponding phase transition. No fine tuning or baroque model building is required. We also calculated the impact on the DM abundance, which is diluted in this scenario as the result of entropy production during reheating. Dilution factors as large as $\Delta \sim 10^3$ are easily attainable in this model, and even larger values are possible in models with a first order phase transition. 

A period of late-time inflation that dilutes the DM abundance can have important implications for a wide range of DM candidates. Among other possibilities, such scenarios naturally accommodate both very light (sub-GeV) and very heavy ($\gsim$10-100 TeV) thermal relics, which are often not viable under standard cosmological assumptions. Ultra-light axions, corresponding to a very high Peccei-Quinn scale, $f_{\rm PQ} \gg 10^{12}$ GeV, are also easily accommodated within this framework, without requiring any anthropic tuning of the misalignment angle. The phenomenology of asymmetric DM models can also be altered in a variety of ways by a period of late-time inflation.  

In general terms, the very stringent constraints from DM direct detection experiments have forced the particle-astrophysics community to focus on WIMPs that possess highly suppressed interactions with nuclei, while still being able to efficiently annihilate in the early universe. Although there are many known ways to accomplish this in model building (coannihilations, resonances, couplings only to Higgs/gauge bosons and/or leptons, etc.), this tension could also be alleviated by a period of late-time inflation. After accounting for the dilution that results from an inflationary event, the revised relic abundance calculation favors DM models with smaller annihilation cross sections, and thus weaker interaction strengths, than would otherwise be expected of a thermal relic. This leads us to anticipate lower DM event rates in direct and indirect detection experiments, as well as at the LHC.

The framework of inflatable dark matter laid out in this paper offers a multitude of opportunities for model building, many of which make connections between the visible and hidden sectors of our universe, or are associated with poorly understood phenomena, such as baryogenesis. As the parameter space of many of our most well-motivated DM candidates becomes more stringently experimentally constrained, scenarios involving a period of late-time inflation will become increasingly attractive.

\begin{appendix}

\section*{Acknowledgments}
We would like to thank Tim Cohen for helpful discussions.  The work of HD is supported by the US Department of Energy under
Grant Contract DE-SC0012704.  The work of DH has been supported by the US Department of Energy under contract DE-FG02-13ER41958. Fermilab is operated by Fermi Research Alliance, LLC, under Contract No. DE-AC02-07CH11359 with the US Department of Energy. SDM is supported by NSF PHY1316617.

\section{Conditions for Inflation}\label{inflation}
The condition $T_s \geq T_{pt}$ is necessary but not sufficient for inflation to occur. In addition, the potential energy stored in the
inflating field must dominate over its kinetic energy: $\dot \phi^2 \leq V_\phi$ (for a review of the physics of inflation, see Ref.~\cite{Baumann:2009ds}).  Here, we briefly describe how this is satisfied in our toy model described in Sec.~\ref{toy}.

The potential from \Eq{thermal potential} can be rewritten as
\beq
\frac{V_\phi}{V_0} =  1 + 2\pL\frac\phi{v_\phi}\pR^2\bL\pL\frac T{T_{pt}}\pR^2-1\bR + \pL\frac\phi{v_\phi}\pR^4 .
\eeq
Since $\phi$ is initially trapped at $\phi=0$, we can make the approximation $V_\phi\sim V_0$, with $V_0 = 3 \mu_\phi^4/(2 \lambda_\phi)$, during inflation.
Before inflation begins, $\phi$ is in thermal equilibrium and hence its kinetic energy can be approximated by $\dot \phi^2 \sim T_s^4$, where $T_s$ is
given by \Eq{T*}. Therefore, we see that as long as $g_s \gsim 3$, which is always the case in this model, the kinetic energy
of $\phi$ is subdominant to $V_\phi$.  Note that the kinetic energy redshifts exponentially as inflation proceeds and thus this condition is satisfied increasingly well as time goes on.

Inflation ends at $T=T_{pt}$ and $\phi$ starts rolling, turning the energy stored in $V_\phi$ into kinetic energy. The equation of motion for $\phi$ is given by the Klein-Gordon equation, $\ddot \phi = -V_\phi'$ (where we ignore Hubble friction because the Hubble scale, $H(T_s) \sim V_0^{1/2}/M_{\rm Planck} \ll \mu_\phi$, is so small). We find
\beq
\ddot \phi = \mu_\phi^2 \phi \bL 1 - \pL \frac T{T_{pt}} \pR^2 - \pL \frac\phi{v_\phi} \pR^2 \bR,
\eeq
which does not have a simple solution for general values of $\phi$. However, soon after the phase transition when $T/T_{pt}=1-\tau$ and $\phi \ll v_\phi$, we may simplify to find $\ddot \phi \sim 2 \mu_\phi^2 \tau \phi$. With the initial condition $\dot \phi \sim 0$, we find the solution $\phi \propto \sinh (t \, \mu_\phi\sqrt{2\tau} )$ which persists until $\phi$ approaches $v_{\phi}$ and the friction term becomes non-negligible. After arriving near the minimum of its potential, the $\phi$ field quickly decays, since for even tiny couplings to the Standard Model ({\it e.g.}, through a Higgs portal term with very small mixing, $\gtrsim 10^{-8}$) the $\phi$ width will be vastly larger than the Hubble scale.

Massless fermions coupled to the scalar field are primarily responsible for maintaining $\vev \phi=0$ during inflation. As the $\phi$ field moves away from zero following its phase transition, these fermions will attain a mass, $m_D = y \, \vev\phi$. Using the condition $\lambda_\phi \lsim \lambda_c$, one can show
\beq
\frac{m_D}{T_{\rm RH}} \gsim 2 \pi \left(\frac{g_s \,g_{\rm RH}}{25 \,N_D^2}\right)^{1/4}.
\label{mD}
\eeq
For all reasonable reheating temperatures considered in our work, we have $g_s \gsim g_{\rm RH} \gsim 10$ and
hence typically $m_D/T_{\rm RH} \gsim \ord{\text{few}}$ by the time $\vev \phi = v_\phi$.  Consequently, these fermions are effectively Boltzmann suppressed following reheating and cannot return $\phi$ to the origin at $\phi=0$, even if $T_{\rm RH} > T_{pt}$.  Thus, the universe promptly re-enters radiation domination and no further inflation will take place.

\end{appendix}

\bibliography{inflatable}

\end{document}